\documentclass[aps,prl,twocolumn,preprintnumbers,amsmath,amssymb,superscriptaddress,floatfix,longbibliography]{revtex4-2}

\usepackage{graphicx}
\usepackage{dcolumn}
\usepackage{bm}

\usepackage{xcolor}
\usepackage{soul}
\usepackage{graphicx}
\usepackage{dcolumn}
\usepackage{bm}
\usepackage{mathrsfs}
\usepackage{amsmath}
\usepackage{amssymb}
\usepackage{amsfonts}

\def\lesssim{\ \raise.3ex\hbox{$<$}\kern-0.8em\lower.7ex\hbox{$\sim$}\ }
\def\gesim{\ \raise.3ex\hbox{$>$}\kern-0.8em\lower.7ex\hbox{$\sim$}\ }

\newcommand{\beginsupplement}{%
        \setcounter{table}{0}
        \renewcommand{\thetable}{S\arabic{table}}%
        \setcounter{figure}{0}
        \renewcommand{\thefigure}{S\arabic{figure}}%
      \setcounter{equation}{0}
        \renewcommand{\theequation}{S.\arabic{equation}}%

     }

\def\tr{{\rm tr}}
\def\Im{{\rm Im}}

\newcommand{\calB}{ \mathcal{B} }
\newcommand{\tmu}{ \tilde{\mu} }
\newcommand{\tmuB}{ \tilde{\mu}_{\rm B} }

\newcommand{\MB}{ M_{\rm B} }
\newcommand{\Kth}{ K_{\rm th} }
\newcommand{\tomega}{ \tilde{\omega} }

\newcommand{\beq}{\begin{equation}}
\newcommand{\eeq}{\end{equation}}

\begin{document}
\title{Tripling Fluctuations and Peaked Sound Speed in Fermionic Matter}

\author{Hiroyuki Tajima}
\affiliation{Department of Physics, Graduate School of Science, The University of Tokyo, Tokyo 113-0033, Japan}
\affiliation{RIKEN Nishina Center, Wako 351-0198, Japan}

\author{Kei Iida}
\affiliation{RIKEN Nishina Center, Wako 351-0198, Japan}
\affiliation{Department of Mathematics and Physics, Kochi University, 780-8520, Japan}

\author{Toru Kojo}
\affiliation{Theory Center, IPNS, High Energy Accelerator Research Organization (KEK), 1-1 Oho, Tsukuba, Ibaraki 305-0801, Japan}

\author{Haozhao Liang}
\affiliation{Department of Physics, Graduate School of Science, The University of Tokyo, Tokyo 113-0033, Japan}
\affiliation{RIKEN Interdisciplinary Theoretical and Mathematical Sciences Program, Wako 351-0198, Japan}

\date{\today}
\begin{abstract}
  A crossover involving three-fermion clusters is relevant to the hadron-quark crossover, which, if occurring in a neutron star, could naturally reproduce the dense-matter equation of state recently deduced from simultaneous observations of neutron-star masses and radii. 
  To understand the crossover mechanism, we examine the role of {\em tripling} fluctuations induced by the formation of three-fermion clusters.
  The phase-shift representation of fluctuations manifests an interplay of bound and scattering states, leading to non-monotonic momentum distributions of baryon-like clusters and peaked sound speed at finite densities. 
  We demonstrate them by applying our approach to a nonrelativistic system of one-dimensional three-color fermions
  analogous to the hadron-quark matter.
\end{abstract}

\maketitle
{\it Introduction.---}
The properties of dense matter above the normal nuclear density, once elusive due to the scarcity of empirical data, are now becoming accessible through recent observational advances.
The long-awaited simultaneous determination of neutron star masses and radii indicates that the equation of state (EOS) of neutron star cores radically stiffens, i.e., pressure $P$ grows rapidly as energy density $\varepsilon$ increases~\cite{Baym_2018,lattimer2021neutron}.
The inferred core density can be high enough for baryons composed of three quarks to overlap.
These together imply that a transition from baryonic matter into dense quark matter, if it indeed occurs as expected, should be smooth.

The quark-hadron continuity/crossover (QHC) scenario, originally focused on the symmetry aspects~\cite{PhysRevLett.82.3956},
was first realized in the EOS through a phenomenological interpolation between hadronic and quark matter~\cite{Masuda:2012kf},
predicting radical stiffening in the crossover domain.
A useful measure of the stiffening is the peak of sound speed, $c_s = (dP/d\varepsilon)^{1/2}$.
The microscopic reasoning, why crossover models not only avoid softening associated with phase transitions but also drive the radical stiffening, is still in debate.
One possible explanation is given by models of quarkyonic matter~\cite{McLerran:2007qj,PhysRevLett.122.122701} in which
the quark Pauli blocking forbids baryons to occupy states at low momenta and therefore make them relativistic, leading to the stiff EOS~\cite{PhysRevLett.132.112701,PhysRevD.104.074005}.
The resulting momentum distribution of baryons is highly non-monotonic.

The previous quarkyonic matter models did not manifestly describe the formation of baryons in dense matter.
In this Letter, we treat three-fermion correlations responsible for the baryon formations
and show how the non-monotonic baryon distribution arises.
Our analyses generalize studies in two-color or isospin QCD in which the QHC have been established by lattice simulations~\cite{10.1093/ptep/ptac137,iida2024two}. 
For these theories, bosons condense at certain chemical potentials, 
and beyond which the transition proceeds smoothly in a similar way as the Bose-Einstein-condensate (BEC) to Bardeen-Cooper-Schrieffer (BCS) crossover~\cite{PhysRev.186.456,leggett1980diatomic,nozieres1985bose,PhysRevLett.71.3202}
(see also reviews~\cite{CHEN20051,zwerger2011bcs,strinati2018bcs,OHASHI2020103739}) experimentally realized in ultracold atoms~\cite{PhysRevLett.92.040403,PhysRevLett.92.120403,PhysRevLett.92.203201} as well as in superconductors~\cite{kasahara2014field,nakagawa2021gate,PhysRevX.12.011016}.
In particular, the lattice simulations confirmed the sound-speed peak~\cite{10.1093/ptep/ptac137} and its qualitative behavior can be captured by simple models treating condensed bosons as mean-fields~\cite{PhysRevD.105.076001,Chiba:2023ftg}.

For three-color QCD, the relevant hadrons are baryons and they must be included as \textit{tripling} fluctuations induced by the formation of three-body bound states. 
The physics of tripling fluctuations is elusive compared to pairing fluctuations in the BEC-BCS crossover but recently there has been some progress.
In cold-atom physics,
a smooth crossover from bound trimers to Cooper triples~\cite{PhysRevA.86.013628} has been discussed in three-\textit{color} (i.e., three hyperfine states) Fermi gases~\cite{PhysRevA.104.023319,PhysRevResearch.4.L012021,PhysRevC.109.055203}.
This scenario is found to be consistent with quantum Monte Carlo (QMC) simulations in a spatially one-dimensional system~\cite{PhysRevA.102.023313}.
Moreover, in one spatial dimension, nonrelativistic three-component Fermi gas with three-body attraction
 accompanies asymptotic freedom and a trace anomaly~\cite{PhysRevLett.120.243002} just like dense QCD matter~\cite{PhysRevLett.129.252702}.
Three-body spectra have been also investigated at finite temperature~\cite{PhysRevResearch.4.L012021}.

In this Letter, we examine the role of tripling fluctuations by generalizing the pairing-fluctuation approach~\cite{nozieres1985bose}.
The phase-shift representation of the multi-body propagator plays a central role in describing the non-monotonic momentum distribution of baryons.
To explore analytic insights we use a model of nonrelativistic three-color fermions in one spatial dimension.
We show how the sound-speed peak is induced by tripling fluctuations in the crossover regime.

{\it Clustering fluctuations.---}
Simply adding many-body cluster contributions causes questions concerning the double counting of constituent particles. 
To handle this problem we first present the thermodynamic potential characterized by the phase shift~\cite{Dashen:1969ep,Blaschke:2013zaa,Lo:2017sde,Andronic:2018qqt}.
There is an important constraint on the asymptotic behaviors of the phase shift and
we find how the contributions from composite states are canceled at high energy.
Such cancellations are crucial to establish the dominance of constituent particles in thermodynamics at high densities.

We are interested in the $N$-body cluster contributions induced by the interaction $V$ to the thermodynamic potential $\Omega$.
Subtracting a term which is already included for the single-particle self-energy,
the $N$-body fluctuations can be written as
\begin{align}
\label{eq:1}
\delta\Omega_N 
 = - T \sum_{K,\omega_\ell}  \tr_N \big[ \ln (1-G_0 V) + G_0 V \big]\,,
\end{align}
where $G_0$ is the bare $N$-body propagator characterized by the total
Matsubara frequency $\omega_\ell=(2\ell+1)\pi T$ with the temperature $T$, the center-of-mass momentum $\bm{K}$,
the relative momenta $\bm{k}_1,\cdots\bm{k}_{N-1}$, and the relative frequencies $\omega_{l_1},\cdots,\omega_{l_{N-1}}$
(where we omitted the argument of $G_0$ for convenience).
The trace is taken for all relative momenta and frequencies as ${\rm tr}_N=\int_{k_1,\cdots,k_{N-1}}$ where $\int_{k}= T\int d \bm{k}/(2\pi)^d\sum_{\omega_l}$.
The Matsubara sum in Eq.~\eqref{eq:1}
can be transformed into the integral representation by deforming the contour as
\begin{align}
\label{eq:2}
\delta \Omega_{N}
& = -T  \sum_{\bm{K}}\int_{-\infty}^{\infty} \frac{d \omega}{\pi} \ln \big( 1+e^{-\omega/T} \big)  \partial_{\omega} \Im \, \Phi 
\,.
\end{align}
where $\Phi={\rm tr}_N\left[ \ln (1-G_0 V ) + G_0 V \right]$.
Here we consider the short-range interaction responsible for cluster formations which can often be written in a separable form.
This allows us to take the trace of $G_0$ in $\Phi$ independently as
$\Phi=\left[ \ln (1-\mathcal{G}_0 V ) + \mathcal{G}_0 V \right]$ with $\mathcal{G}_0={\rm tr}_N[G_0]$.

Now we introduce the phase $\varphi$ of propagators as $\mathcal{G}/\mathcal{G}_0 = |\mathcal{G}/\mathcal{G}_0| e^{i \varphi}$ with the trace $\mathcal{G}$ of
the dressed $N$-body propagator $G=G_0+G_0VG$.
$\varphi$ is subject to a constraint that interactions cannot change the size of the state space.
Any propagators must satisfy the sum rule
\begin{align}
\label{eq:3}
\int_{-\infty}^\infty d \omega \,\Im \big( \mathcal{G} - \mathcal{G}_0 \big)
= -\int_{-\infty}^\infty d \omega\, \partial_\omega \ln (\mathcal{G}/\mathcal{G}_0)=0 \,,
\end{align}
which leads to
$\varphi (\bm{K},\infty) = \varphi (\bm{K},-\infty) = 0$,
with setting the phase at $\omega = -\infty$ to zero. 
Since we have
$\Im \, \Phi =
- \varphi + |\mathcal{G}_0/\mathcal{G} | \sin \varphi
$,
Eq.~\eqref{eq:3} leads to
\begin{align}
\Im \, \Phi (\bm{K},\omega \rightarrow \pm \infty) = 0 \,.
\end{align}
%
These boundary conditions have important implications.
The phase $\varphi$
increases by $\pi$ when $\omega$ passes the bound-state pole (and also the resonances). 
However, the increased phase eventually must be reduced back to the original value because of $\varphi (\bm{K},\infty) = \varphi (\bm{K},-\infty) = 0$. 
This indicates that $\partial_\omega \Im \,\Phi(\bm{K},\omega) $ must be negative in some interval of $\omega$ beyond the continuum threshold,
and this negative contribution tends to cancel the positive bound-state contributions.

The convoluted integral of the phase with the thermal factor $\ln (1+e^{-\omega/T})$ in Eq.~\eqref{eq:2} preferentially picks up the bound-state contributions at low temperatures, 
while at high temperature the scattering states also contribute and tend to cancel the bound-state contributions.
Hence, in the high-temperature limit the thermodynamics should be dominated by single-particle contributions as expected.
Such a mechanism plays a crucial role in understanding the microscopic mechanism of the hadron-quark crossover accompanied by tripling fluctuations associated with baryon formations.
In this work we examine the occupation probability of baryonic states at finite density, and show how the above cancellation works differently for different $\bm{K}$.


{\it Demonstration of tripling fluctuations.---}
To see the role of clustering fluctuations beyond the two-body level,
we consider a nonrelativistic three-color fermions with the color-singlet three-body attractive interaction in one spatial dimension,
of which the dynamics is described by the Hamiltonian~\cite{PhysRevLett.120.243002}
\begin{align}
H&=\sum_{p,\alpha}\xi_{p}c_{{p},\alpha}^\dag c_{{p},\alpha}
+V\sum_{P} B^\dag(P) B(P)\,,
\end{align}
where $\xi_{p}=p^2/(2m)-\mu$ is the kinetic energy of a fermion with mass $m$, measured from the chemical potential $\mu$,
$c_{p,\alpha}$ is the annihilation operator with color $\alpha={\rm r}, {\rm g}, {\rm b}$, and
$B(P)= \sum_{q,k} c_{\frac{P}{3}-k+\frac{q}{2},{\rm r}}c_{\frac{P}{3}+k+\frac{q}{2},{\rm g}}c_{\frac{P}{3}-q,{\rm b}}$ 
 is the three-body operator.
Here
we consider the population-balanced case (without other degrees of freedom such as spins and flavors).
The contact-type three-body coupling $V$ is associated with the three-body binding energy $\mathcal{B}$ as~\cite{PhysRevLett.120.243002}
\begin{align}
    \frac{1}{V(\Lambda) }=-\frac{m}{2\sqrt{3}\pi}\ln\left(\frac{ \calB + \Lambda^2/m }{ \calB }\right)\,,
\end{align}
where  $\Lambda$ is the momentum cutoff and the coupling $V(\Lambda)$ is renormalized with $\calB \, (> 0)$ fixed.
The emergence of nonzero $\mathcal{B}$ is the consequence of quantum anomaly~\cite{PhysRevLett.120.243002}, 
and is expected to induce strong tripling fluctuations at finite densities.

Using Eq.~\eqref{eq:1}
we obtain the
thermodynamic potential 
as
$\Omega=\Omega_{\rm HF}+\delta\Omega_{3}$,
where
    $\Omega_{\rm HF}=-3T\sum_{k}
    \ln\left(1+e^{-\xi_{k}^{\rm HF}/T}\right)$
with $\xi_{k}^{\rm HF}=\xi_k+\Sigma_{\rm HF}$
is the Hartree-Fock~(HF) contribution.
$\Sigma_{\rm HF}=V\left({n_0}/{3}\right)^2$    
is the HF self-energy
with the non-interacting fermion number density $n_0=3\sum_{k}f(\xi_k)$ 
and $f(x)=1/(e^{x/T}+1)$ is
the Fermi distribution function.
The tripling fluctuation term is
\begin{align}
\label{eq:8v2}
    \delta\Omega_{3}
    =-T\sum_{K,i\omega_\ell}
    \Bigl[
    \ln\left( 1-V\mathcal{G}_0 \right)
    +V\mathcal{G}_0
    \Bigr]\,,
\end{align}
where $\mathcal{G}_0 \equiv \mathcal{G}_0(K,i\omega_\ell)$ is the bare three-body propagator whose internal momenta and frequencies are traced out.
Performing the analytical continuation of $\mathcal{G}_0$ to the real frequency $\omega+i\delta$ with infinitesimally small number $\delta$, we obtain~\cite{PhysRevResearch.4.L012021}
\begin{align}
\label{eq:xi}
    \mathcal{G}_0(K, \omega )
    &=
    \sum_{k,q}
    \frac{ (1-f_1) (1-f_2) (1-f_{3}) + f_{1} f_{2} f_{3} }
    {\omega+i\delta-\xi_{1}-\xi_{2}-\xi_{3}}
    \,,
\end{align}
where $f_i=f(\xi_{i})$ with
single-particle energies
$\xi_{1}=
\xi_{\frac{q}{2}+\frac{K}{3}+k}^{\rm HF}$,
$\xi_2=\xi_{\frac{K}{3}-q}^{\rm HF}$,
and
$\xi_3=\xi_{\frac{q}{2}+\frac{K}{3}-k}^{\rm HF}$.
$\Lambda$ is applied to the momentum summations in Eq.~\eqref{eq:xi}.

\begin{figure}[t]
    \centering
    \includegraphics[width=7.5cm]{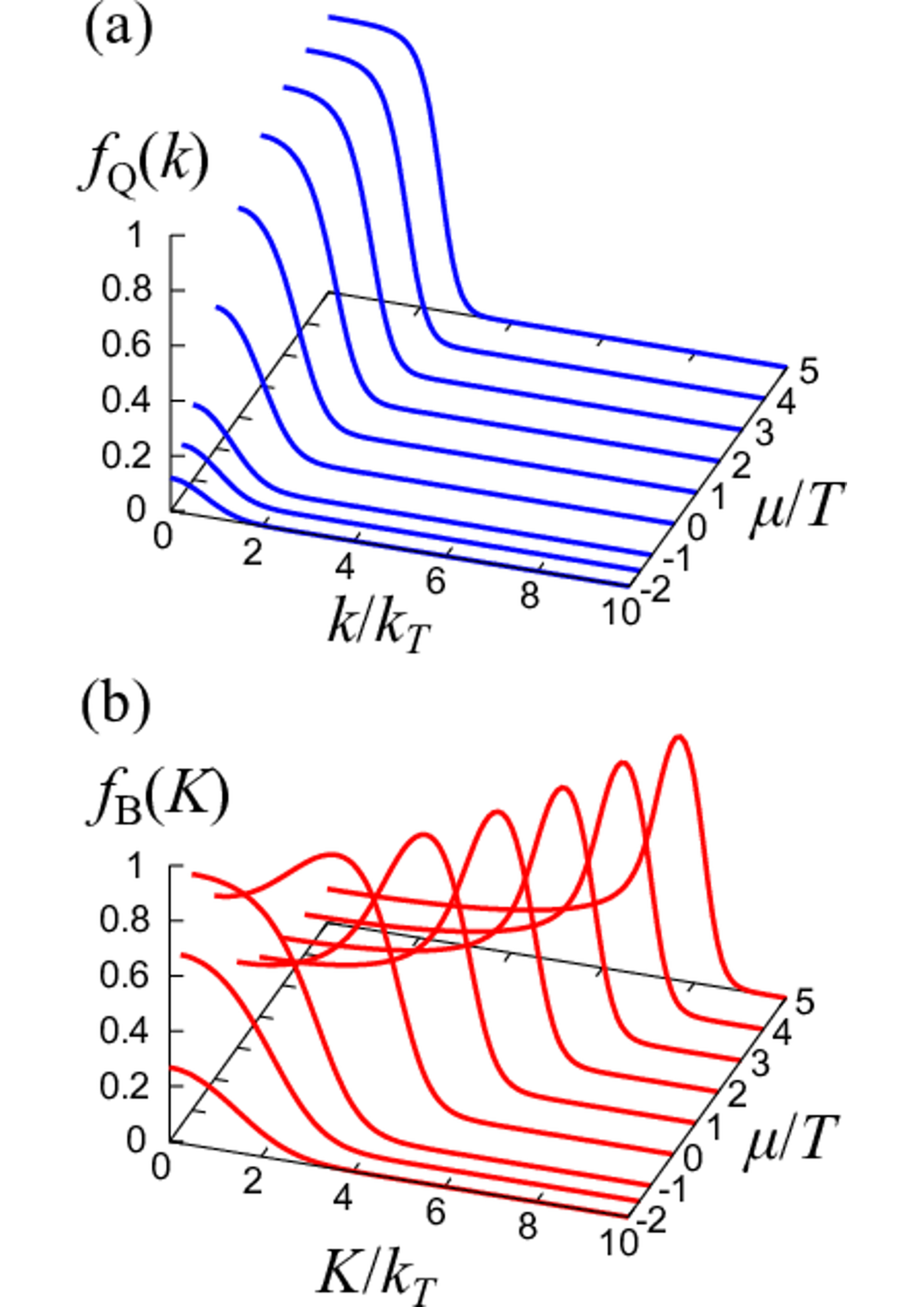}
    \caption{The momentum distribution functions of (a) constituent (quark-like) fermions, $f_{\rm Q}(k)$, and (b) (baryon-like) three-body states, $f_{\rm B}(K)$, at $T/\calB=0.2$. $k_T=\sqrt{2mT}$ is the momentum scale associated with the temperature $T$.}
    \label{fig:2}
\end{figure}

Below we set $f_i \rightarrow 0$ by neglecting the structural changes of a baryon caused by the statistical restriction on the intermediate quark states.
This approximation is valid when $K$ is sufficiently large so that quarks become free from the Pauli blocking effects.
At low $K$, this approximation is not necessarily valid, but the contributions from low $K$ turn out to be suppressed anyway by the phase-shift effects.
Then the integral in Eq.~\eqref{eq:xi} can be carried out analytically as
\begin{align}
\label{eq:xi2}
    \mathcal{G}_0(K,\omega)
    \simeq
    -\frac{m}{2\sqrt{3}\pi}
    \ln\left(\frac{ - \tomega -i\delta + \Lambda^2/m }{ - \tomega -i\delta }\right)\,,
\end{align}
where we introduced
$\tomega = \omega + \tmuB - E_{\rm B}^{\rm kin}(K)$
with $E_{\rm B}^{\rm kin}(K) = {K^2}/2M_{\rm B}\equiv K^2/(6m)$ and
$\tmuB =3\tmu\equiv 3(\mu-\Sigma_{\rm HF})$.
After setting $\Lambda \rightarrow \infty$ in $V$ and $\mathcal{G}_0$,
we obtain the phase shift of the three-body propagator as
\begin{align}
\label{eq:10}
\varphi (\tomega)
=
\pi \Theta( \tomega + \calB ) \Theta( - \tomega)
+ \Theta (\tomega) \varphi_{\rm scatt} (\tomega) \,,
\end{align}
where $\Theta(x)$ is the step function.
In Eq.~\eqref{eq:10},
the first term is the bound-state contribution and
\begin{align}
\varphi_{\rm scatt}(\tomega) = \tan^{-1} \left[\frac{ \pi }{ \ln (|\tomega | /\calB ) }\right] \,,
\end{align}
is the scattering-state contribution.
$\varphi_{\rm scatt}(\tomega)$
approaches $\pi$ for $\tomega \rightarrow 0_+$; passes $\pi/2$ for $\tomega \rightarrow +\calB$;
and approaches zero for $\tomega \rightarrow +\infty$.
Accordingly,
$\varphi (\tomega)$ satisfies the boundary condition $\varphi(-\infty) = \varphi(\infty)=0$.
\begin{figure}[t]
\vspace{-0.8cm}
    \centering
    \includegraphics[width=8.6cm]{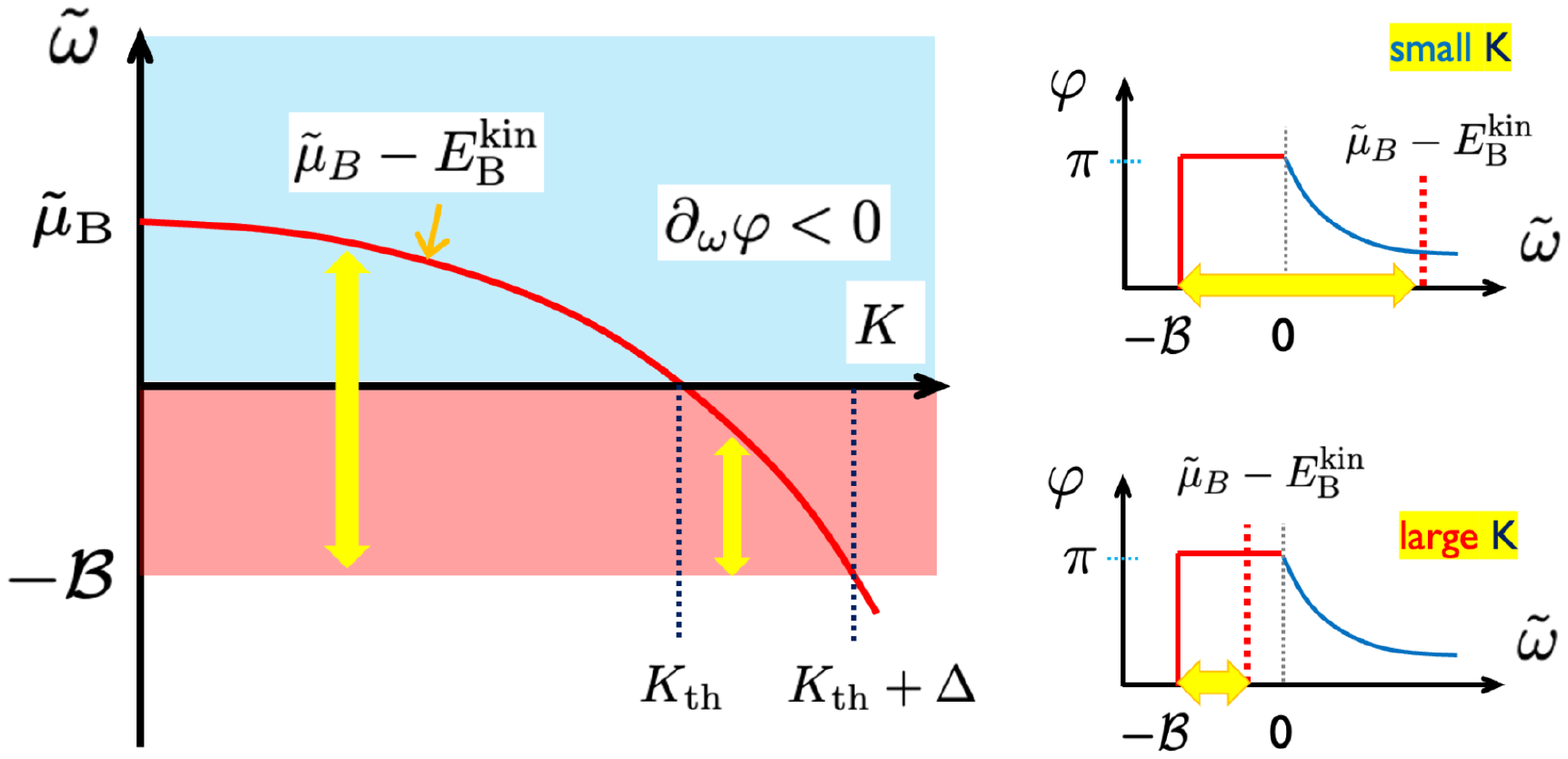}
\vspace{-1.5cm}    
    \caption{Schematics of the phase shift $\varphi$ at various $\omega$ and $K$. 
    For a greater $K$, the domain with $\partial_\omega \varphi <0$ (i.e., scattering state) becomes narrower so that the bound-state contributions on thermodynamic quantities survive from the cancellation with the scattering-state contributions.
    For a smaller $K$, the scattering-state contributions more strongly cancel the bound-state contributions so that the baryonic momentum distribution $f_{\rm B} (K)$ is smaller as we shall show below.
    }
    \label{fig:1}
\end{figure}
Taking the derivative leads to
\begin{align}
\frac{1}{\pi} \partial_{\tomega} \varphi (\tomega)
=
\delta( \tomega + \calB ) 
- \Theta (\tomega)  \frac{1}{\, \tomega \,}  \frac{1}{\,  \ln^2 (\tomega /\calB )  + \pi^2 \,} \,.
\end{align}
The sign of the scattering contribution is opposite to the bound state one, as mentioned earlier, leading to their cancellation (see also Fig.~\ref{fig:1}).
Substituting this expression to $\delta\Omega_3$ reads
\begin{align}
\delta\Omega_{3} 
= &- T \sum_K  \ln \big[ 1+e^{-(  - \calB +  E_{\rm B}^{\rm kin}  - \tmuB  )/T } \big] 
\notag \\
&+T \sum_K \int_0^{\infty} \frac{ d\omega }{\omega} 
\frac{ \ln \big[ 1+e^ {- ( \omega + E_{\rm B}^{\rm kin} - \tmuB )/T }  \big]  }{  \ln^2 ( \omega / \calB )  + \pi^2  } \,.
\end{align}
%
%
%
Using the thermodynamic identity $n=-{\partial \Omega}/{\partial\mu}$, where $n$ is the fermion number density,
we write
$n=n_{\rm HF}+n_{\rm fluc}$ with
$n_{\rm HF} = - \partial \Omega_{\rm HF}/\partial \mu$ and $n_{\rm fluc} = - \partial \delta\Omega_{3}/\partial \mu$ given by
\begin{align}
    n_{\rm HF}=3\sum_{k} f_{\rm Q} (k) \,,
    ~~~~
    n_{\rm fluc}=3 \sum_{K} f_{\rm B} (K) \,.
\label{eq:nHF0}
\end{align}
The first term is the HF contribution as $f_{\rm Q}(k) \equiv f (\xi_k^{\rm HF} )$ and the second one is the baryon-like fluctuation contributions given by
\begin{align}
\label{eq:16}
f_{\rm B}(K)
    & =f( - \calB + E_{\rm B}^{\rm kin} - \tmuB )\cr
    & \ \ \ -
    \int_{0}^{\infty}\frac{d\omega}{\omega}
        \frac{ f(\omega +  E_{\rm B}^{\rm kin} - \tmuB) }{ \ln^2( \omega/\calB)+\pi^2}\,,
\end{align}
where the qualitative behavior of $f_{\rm B}(K)$ is governed by the cancellation of bound and scattering state contributions as shown in Fig.~\ref{fig:1}.
The numerical results for $f_{\rm Q}(k)$ and $f_{\rm B}(K)$ as functions of the momenta and $\mu/T$ are shown in Fig.~\ref{fig:2}.
One can find a
step-like distribution of $f_{\rm Q}(k)$ and
a momentum-shell structure of $f_{\rm B}(K)$ at larger $\mu/T$, which is consistent with the duality model~\cite{PhysRevLett.132.112701}.
The numerical results of $n$ are also shown in the Supplement and qualitatively agree with the QMC result~\cite{PhysRevA.102.023313}.

To explore the analytic insights, we consider the low temperature limit ($T\rightarrow 0$).
%
In the dilute limit such that $ - \calB \le \tmuB \le 0$, the single-particle contributions vanish while baryonic contributions read
%
 $f_{\rm B}^{T= 0} (K) \big|_{\rm dilute}
    = \Theta (\tmuB + \calB - E_{\rm B}^{\rm kin} )$,
where states are fully occupied up to $\sqrt{2\MB(\tmuB + \calB)}$.
On the other hand, when $\tmuB \ge 0$, the single particle contributions become nonzero
and at the same time the scattering channel in baryonic contributions gets open.
Here we introduce the Fermi momentum $k_{\rm F}=\sqrt{2m\tmu}$ of quark-like fermions
and the threshold momentum $K_{\rm th} \equiv \sqrt{2\MB \tmuB} = 3 k_{\rm F}$  where the continuum for the baryonic channel is open.
%
%
%
Using these, the quark distribution is $f_{\rm Q}^{T=0} (k) = \Theta (k_{\rm F}-k)$
and the baryon distribution is analytically obtained from Eq.~\eqref{eq:16} as 
\begin{align}
 f_{\rm B}^{T= 0} (K)
    &= \Theta ( \Kth + \Delta - K ) \Theta (K- \Kth )
    \notag \\
    & \hspace{-1.2cm}
    + \Theta (  \Kth -K ) \bigg( 
    \frac{1}{2} - \frac{1}{\pi} \tan^{-1} \bigg[ \frac{ \ln \frac{ \Kth^2 - K^2 }{ 2 \MB \calB } }{\pi} \bigg] 
    \bigg) \,.
\end{align}
where we also introduced a function characterizing the thickness of the momentum shell
\begin{align}
\Delta (\Kth) \equiv \sqrt{ \Kth^2 + 2\MB \calB \,} - \Kth \,.
\end{align}
When $K$ approaches $\Kth$,
the distribution approaches the maximum, $ f_{\rm B}^{T= 0} (K\rightarrow \Kth) \rightarrow 1$;
In dense limit, $ \Kth^2  \gg 2 \MB \calB$, the distribution at $K=0$ vanishes as $ f_{\rm B}^{T= 0} (0) \rightarrow 0$.

In terms of the McLerran-Reddy model for quarkyonic matter~\cite{PhysRevLett.122.122701},
this result can be interpreted as the formation of the quark Fermi sphere up to $k=k_{\rm F}$ and the momentum-shell formation of baryonic excitations.
While in the duality model~\cite{PhysRevLett.132.112701} the suppression of $f_{\rm B}(K<3k_{\rm F})$ is associated with the Pauli blocking effect of quarks,
this suppression in the present case is caused by the negative scattering contributions that
are necessary to cancel the doubly-counted quark contributions in baryonic terms.
Since our methodology manifestly treats three quarks in a baryonic state, the quark Pauli principle in the duality model is expected to be automatically taken into account.

{\it Sound speed---}
Let us turn to the squared isothermal sound speed $c_s^2$. 
In the nonrelativistic system, 
it can be rewritten as $c_s^2=({n}/{m})\left({\partial n}/{\partial \mu}\right)_T^{-1}$,
which becomes equivalent to the adiabatic sound speed at $T=0$.
We normalize $c_s^2$ by the Fermi velocity $v_{\rm F}=\pi n/3m$.
\begin{figure}[t]
    \centering
	\includegraphics[width=8cm]{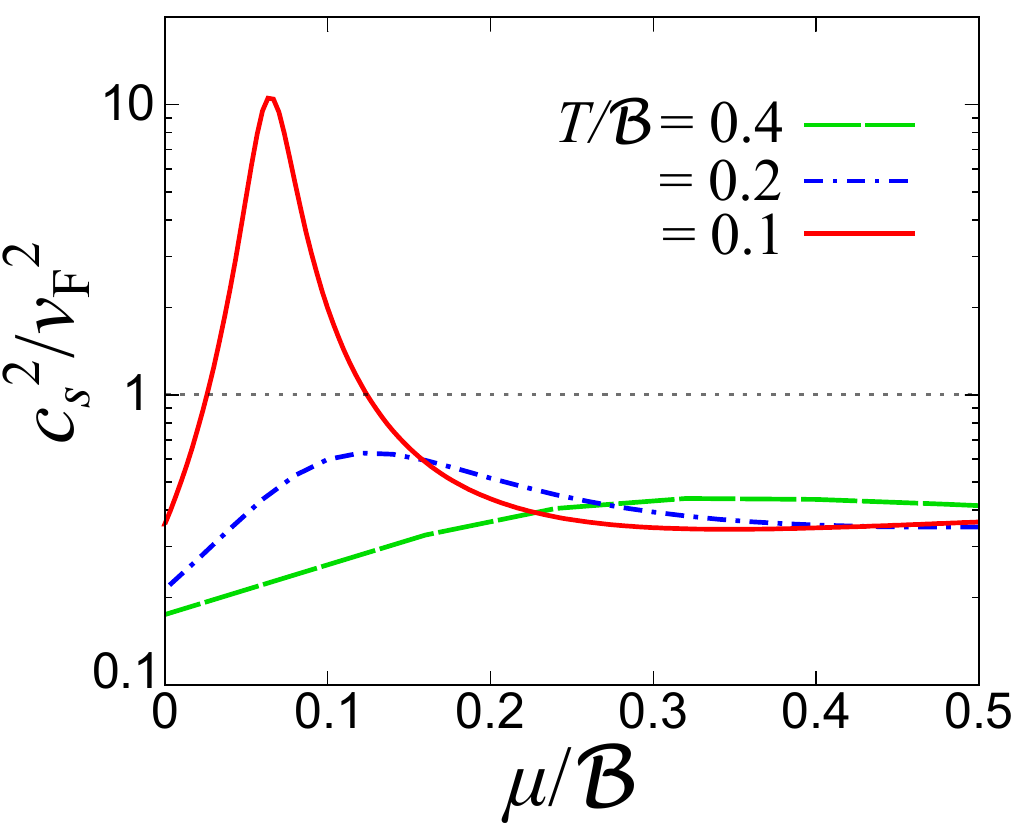}
    \caption{Squared isothermal sound speed $c_s^2/v_{\rm F}^2$ normalized by the Fermi velocity $v_{\rm F}=\frac{\pi n}{3m}$ at different $T$. The horizontal dotted line represents the nonrelativistic conformal limit $c_s^2/v_{\rm F}^2=1$.}
    \label{fig:3}
\end{figure}
One can see the peaked behavior of $c_s^2/v_{\rm F}^2$ in the evolution of $\mu$ in Fig.~\ref{fig:3} at different $T/\mathcal{B}$.
At high densities, $c_{s}^2/v_{\rm F}^2$ slowly approaches unity from below.

To understand the peaked behavior of $c_s^2$, the most essential is the density susceptibility $\chi=\left({\partial n}/{\partial\mu}\right)_T\equiv\chi_{\rm HF}+\chi_{\rm fluc}$, where
$\chi_{\rm HF}=\left({\partial n_{\rm HF}}/{\partial\mu}\right)_T$ and
$\chi_{\rm fluc}=\left({\partial n_{\rm fluc}}/{\partial\mu}\right)_T$ are the HF and fluctuation contributions, respectively.
%
%
While $\chi_{\rm HF}$ is always positive,
$\chi_{\rm fluc}$ can be negative
due to the scattering-state contribution (see also Supplement).
Intuitively, the strong suppression of $f_{\rm B}(K)$ at small $K$ in Fig.~\ref{fig:2}(b) indicates the decrease of $n_{\rm fluc}$ with increasing $\mu$, that is, $\left({\partial n_{\rm fluc}}/{\partial \mu}\right)_T<0$.
The cancellation of $\chi_{\rm HF}$ and $\chi_{\rm fluc}$ leads to the enhancement of $c_{s}^2\propto\chi^{-1}$.
In this way,
one can understand physics of clustering fluctuations in the crossover involving the peaked sound speed,
which allows for a schematic (but based on microscopic many-body physics) viewpoint on the quarkyonic-matter equation of state as  shown in Supplement.

{\it Summary---}
We have investigated the role of tripling fluctuations in the equilibrium properties of nonrelativistic three-color fermions in one spatial dimension, which can be regarded as an analog quantum simulator for the quarkyonic-type hadron-quark crossover in dense matter and is possibly accessible in future cold-atom experiments. 
Using the phase-shift approach to describe tripling fluctuations,
we have elucidated how the momentum distributions of one-body and three-body fermion states
change at finite densities.
Our simplified approach is a minimal model that not only reproduces the duality model for quarkyonic matter, but also
is qualitatively consistent with the QMC results.
For future perspective, it is interesting to apply our approach to systems closer to real QCD matter.

\begin{acknowledgements}
The authors thank Joaqu\'{i}n E. Drut, Yaqi Hou, and Eiji Nakano for useful discussions.
This research was supported in 
part by Grants-in-Aid for Scientific Research provided by 
JSPS through Grants No.~JP18H05406, No.~JP22H01158, No.~JP22K13981, No.~JP23H01167, and No.~JP23K25864.
\end{acknowledgements}

\bibliographystyle{apsrev4-1}
\bibliography{reference.bib}

\clearpage

\beginsupplement

\setcounter{equation}{0}
\setcounter{figure}{0}
\setcounter{table}{0}
\setcounter{page}{1}
\makeatletter
\renewcommand{\theequation}{S\arabic{equation}}
\renewcommand{\thefigure}{S\arabic{figure}}
\renewcommand{\bibnumfmt}[1]{[S#1]}

\section{Equation of state and phase shift}

\begin{figure}[t]
\vspace{-0.8cm}
    \centering
    \includegraphics[width=8.6cm]{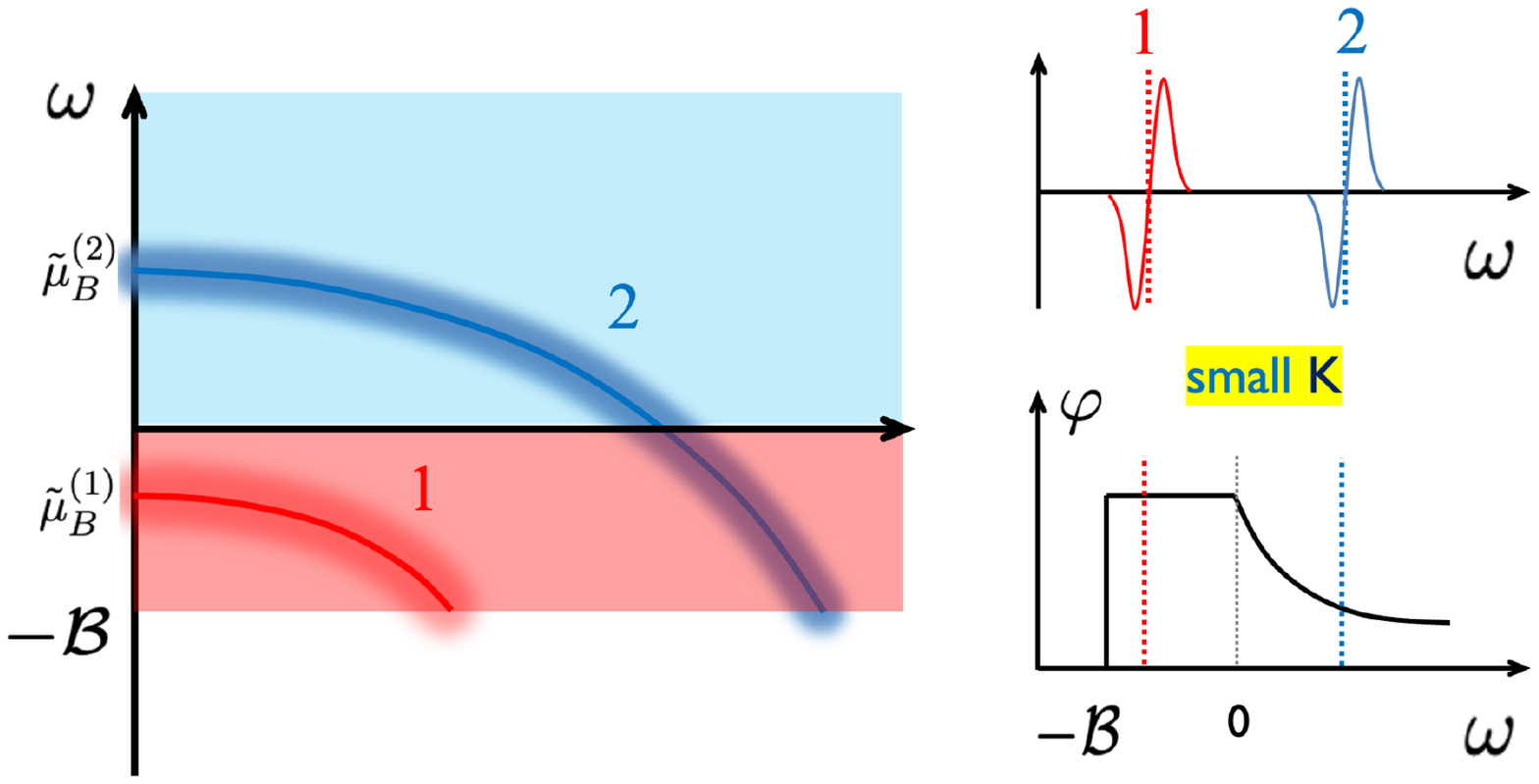}
\vspace{-1.5cm}    
    \caption{ Schematics for elements contributing to the density susceptibility $\chi$. The second derivatives $\partial^2 f ({\omega} + E_{\rm B}^{\rm kin} -\tmuB)/\partial \mu^2$ (upper right corner)
    are convoluted with $\varphi ({\omega})$ for several $\tmuB$. The scattering region gives the negative contributions to $\chi$.
    }
    \label{fig:s1}
\end{figure}

We study equations of state (EOS) and examine the consequence of clustering fluctuations in terms of the phase-shift representation.
First, we note that at any nonzero temperature (except for the region near the quantum critical point $\mu=0$) the EOS shows crossover behaviors, provided that the phase $\varphi$ changes continuously during the variation of $\mu$.
In the present analyses, we ignore the structural changes in baryons as also assumed in the duality model~\cite{PhysRevLett.132.112701}, and hence
$\varphi$ is taken the same as the vacuum case.
For both the HF and baryonic parts, the derivatives of $\Omega$ with respect to $\mu$ can be converted into the form
\beq
\frac{\partial^n \Omega}{ \partial \mu^n }
= - T\sum_{K}\int_{-\infty}^{\infty}\frac{d\omega}{\pi}
\varphi (\omega)
\frac{ \partial^n  }{ \partial \mu^n } 
f(\omega + E_{\rm B}^{\rm kin} - \tilde{\mu}_{\rm B})
\,.
\eeq
As far as we take $T$ nonzero, the derivatives of the Fermi-Dirac distributions do not yield any singularities for $\omega \in {\bf R}$.
Although $\varphi(\omega)$ changes discontinuously when $\omega$ passes every threshold of a new excitation, it does not contain any singularities.
Hence $\partial^n \Omega/\partial \mu^n$ for the present EOS are continuous with respect to changes in $\mu$, meaning that the EOS is the crossover type.

Although the derivatives of $\Omega_{\rm HF}$ and $\delta\Omega_{3}$ are both continuous, 
they are not separately satisfy the thermodynamic stability condition, $\chi = - \partial^2 \Omega/\partial \mu^2 >0$. 
For $N$-body correlation terms, the density susceptibility $\chi_{\rm fluc} = - \partial^2 \delta\Omega_{3}/\partial \mu^2$ becomes negative when $\mu$ picks up the
scattering-state contributions with $\partial \varphi/\partial \omega <0$.
To see this it is useful to note
\beq
\frac{ \partial^2 f(x-\mu) }{ \partial \mu^2 }
= T^{-2} \frac{\, e^{(x-\mu)/T} - e^{- (x-\mu)/T} \,}{ (e^{(x-\mu)/T}+1)^2 (e^{-(x-\mu)/T}+1)^2 } \,.
\eeq
The function is localized and changes the sign around $x=\mu$, as shown in Fig.~\ref{fig:s1}.
If we convolute this function with increasing (decreasing) functions, the integral over $\omega$ becomes positive (negative).

As we see shortly, for large $\calB$ and low temperature, our approximate calculations yield the negative susceptibility in total near $\mu=0$, i.e., 
the scattering-state contributions are too large.
This is purely an artifact of our approximate calculations.
Unless stated otherwise we display the results which avoid the dangerous domains.

\section{Fermion number density}
\label{sec:3}
\begin{figure}[t]
    \centering
    \includegraphics[width=7.5cm]{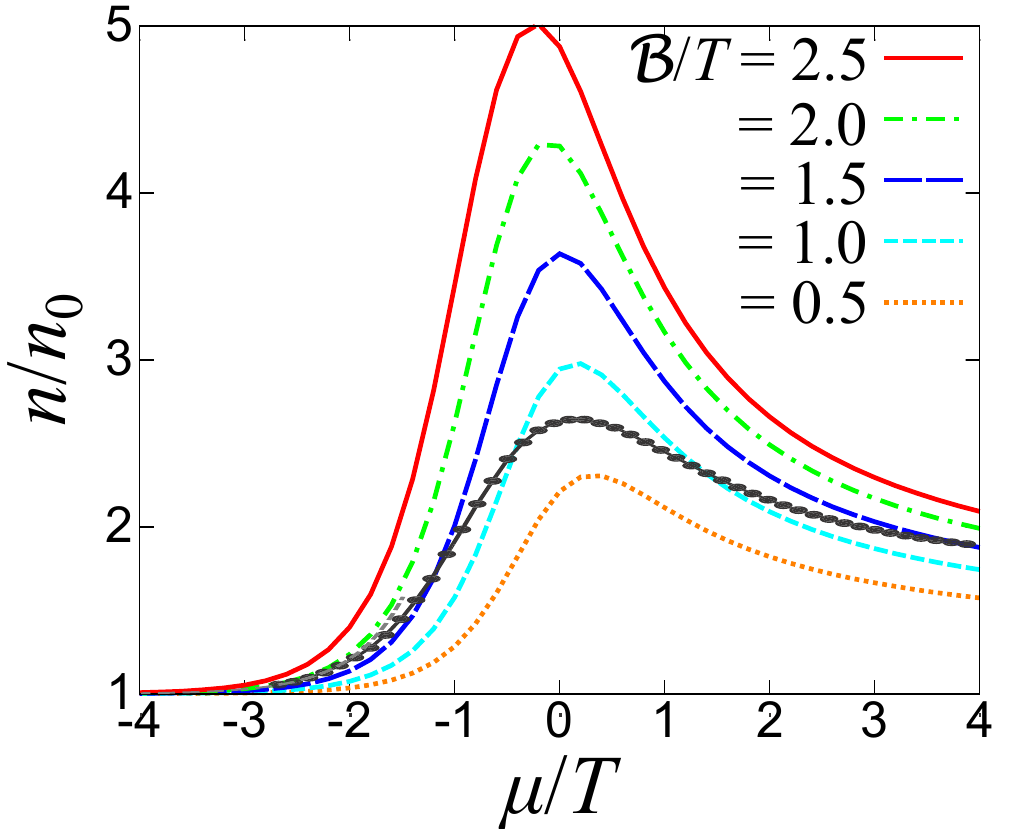}
    \caption{Fermion number density $n/n_0$ as a function of $\mu/T$. $n_0$ is the non-interacting density.
    For comparison, we also show the QMC results (filled circles) and the third-order virial expansion (gray dashed line) at $\calB/T=2$~\cite{PhysRevA.102.023313}.
    }
    \label{fig:s2}
\end{figure}

Here we proceed to display the numerical results of the fermon number density in our model study and compare them with the QMC results~\cite{PhysRevA.102.023313}.
Shown in Fig.~\ref{fig:s2} is the normalized number density $n/n_0$ as a function of $\mu/T$. 
As a baseline, we consider the ideal-gas number density $n_0 = 3\sum_k f(\xi_k)$.

The behavior of $n/n_0$ can be understood as follows: 

(i) Our result in the dilute regime below the single particle threshold ($\mu/T<0$) agrees well with the QMC result.
More explicitly, our approach can reproduce the virial expansion up to the third order of the fugacity $z=e^{\mu/T}$~\cite{PhysRevLett.120.243002} as
\begin{align}
\label{eq:virial1}
    n_{\rm fluc}
        &= \frac{9}{\lambda_T}z^3
    \delta b_3 +O(z^4),
\end{align}
with
\begin{align}
\label{eq:virial2}
    \delta b_3=
    \frac{1}{\sqrt{3}}
    \left[e^{\calB/T}
    -\int_{0}^{\infty}\frac{d\omega}{\omega}\frac{e^{-\omega/T}}{\ln^2(\calB/\omega)+\pi^2}\right],
\end{align}
where $\lambda_T=\sqrt{\frac{2\pi}{mT}}$ is the thermal de Broglie length.

(ii) Beyond the dilute regime, both the QMC and our results show the rapid growth of $n/n_0$ toward $\mu/T = 0$.
This clearly shows the importance of the tripling fluctuations.
In particular, since a three-body bound state has heavy mass $M_{\rm B}=3m$, even small change in $\mu$ can increase its density significantly;
the shift $\mu \rightarrow \mu + \Delta \mu$ is balanced with the energy increase $K^2/2M_{\rm B} \rightarrow (K+\Delta K)^2/2M_{\rm B}$,
allowing large increase in $K$.
This mechanism should be common for the QMC and our computations.
Concerning the quantitative difference, the likely reason is that
our computations, which neglect the structural changes or dissociation of baryons,
overestimate baryonic contributions;
if the Pauli blocking, which is more important at lower $K$, 
cuts off attractive correlations, 
then baryons at low $K$ dissociates toward denser regime (or the $\varphi=\pi$ domain closes),
tempering the growth of $n/n_0$.
The studies of this dissociation effects are left for the future.

(iii) The rapid growth of baryonic contributions is quenched when the scattering channel sets in.
For $\mu/T $ greater than $0$, the baryonic correlations cancel among themselves 
and the thermodynamics is dominated by single particle energies. 
As a result $n/n_0$ approaches $1$.

\section{Schematic description of peaked speed sound in quarkyonic matter}

To characterize thermodynamic properties obtained by the phase-shift approach for clustering fluctuations, we develop the following schematic description for quarkyonic-matter EOS.
We begin with the comparison between the baryonic and quark EOS at a sufficiently low temperature
keeping mass and kinetic terms. 
In this section, we measure the energy density explicitly including the mass terms.
In the baryonic case the energy density reads
\beq
\varepsilon_{\rm B} (n_{\rm B}) = (M_{\rm B} -\calB) n_{\rm B} + c \frac{ n_{\rm B}^3}{M_{\rm B}} \,,
\eeq
while in the quark case
\beq
\varepsilon_{\rm Q} (n_{\rm B}) = N_c \bigg( m n_{\rm B} + c \frac{ n_{\rm B}^3}{m} \bigg) \, ,
\eeq
where $c$ is a numerical constant associated with the kinetic term
and
$N_{c} (=3)$ is the color degree of freedom.

If we neglect the kinetic energies, $\varepsilon_{\rm B}$ and $\varepsilon_{\rm Q}$ are comparable with each other.
In dilute regime quarks contribute to the EOS only through the mass of baryons,
while in dense regime quarks directly contribute to the EOS.
This similarity completely breaks down for the pressure.
We compute the pressure $P=n_{\rm B}^2 \partial (\varepsilon/n_{\rm B})/\partial n_{\rm B}$ 
which measures the density evolution of the energy per particle. 
For each case one can estimate
\beq
P_{\rm B} = 2c \frac{ n_{\rm B}^2 }{M_{\rm B}} \simeq \frac{\, 2c \,}{N_c} \frac{ n_{\rm B}^2 }{ m}\,,~~~~~~P_{\rm Q} = 2c N_c \frac{ n_{\rm B}^2 }{m}\,.
\eeq
where we take $M_{\rm B} \simeq N_{c}m$.
The pressure dominated by quarks are larger by the factor $O(N_c^2)$ if we compare them at the same $n_{\rm B}$.

If we directly compare $\varepsilon_{\rm B}$ and $\varepsilon_{\rm Q}$ at the same $n_{\rm B}$,
the binding energy and smaller kinetic energy of baryonic models would let us choose the baryonic matter as the ground state.
However, as we have examined in detail, baryonic correlations should be canceled by the scattering-state contributions when the scattering channels are open.
Schematically we can mimic such effects on EOS through the following parametrization
\beq
\varepsilon (n_{\rm B}) = (1-w) \varepsilon_{\rm B} +  w \varepsilon_{\rm Q} \,,
\eeq
where $w$ is a monotonically increasing weight function with the asymptotic behavior $w\rightarrow 0$ for $n_{\rm B}\rightarrow 0$ and $w\rightarrow 1$ for $n_{\rm B}\rightarrow \infty$.
The chemical potential reads
\beq
\mu_{\rm B} = \frac{\partial \varepsilon_{\rm B} }{\partial n_{\rm B}} 
+ \frac{\partial (- \varepsilon_{\rm B} + \varepsilon_{\rm Q} ) }{\partial n_{\rm B}}  w
+( - \varepsilon_{\rm B} +  \varepsilon_{\rm Q} ) \frac{\partial w}{\partial n_{\rm B}} 
\,,
\eeq
where $w$ and $\partial w/\partial n_{\rm B}$ as well as their coefficients are all positive,
meaning that $\mu_{\rm B}$ is enhanced from that of the purely baryonic matter.
Such enhancement is regarded as modest because the leading term $\sim N_c m$ is much larger than the binding and kinetic energies.

As we discussed in the main text, the most essential part of the peaked sound speed is the density susceptibility $\chi$.
Its inverse is given by
\begin{align}
\label{eq:chi_inv}
\chi^{-1} 
&= \frac{\partial^2 \varepsilon_{\rm B} }{(\partial n_{\rm B})^2} 
+ \frac{\partial^2 (- \varepsilon_{\rm B} + \varepsilon_{\rm Q} ) }{ (\partial n_{\rm B})^2 }  w
\notag \\
&+ 2 \frac{\partial (- \varepsilon_{\rm B} + \varepsilon_{\rm Q} ) }{ \partial n_{\rm B} }  \frac{\partial w}{\partial n_{\rm B}} 
+( - \varepsilon_{\rm B} +  \varepsilon_{\rm Q} ) \frac{\partial^2 w}{ (\partial n_{\rm B})^2 }  
\,.
\end{align}
In the r.h.s. of Eq.~\eqref{eq:chi_inv}, the first three terms are all positive.
However, the last term can be negative since $\partial^2 w/(\partial n_{\rm B})^2$ must be negative at least in some interval of $n_{\rm B}$,
otherwise it is not possible to satisfy the asymptotic behaviors for $n_{\rm B}\rightarrow 0$ and $n_{\rm B} \rightarrow \infty$.
Too radical change from baryonic to quark matter makes the susceptibility negative,
violating the thermodynamic stability condition.
On the other hand, if the transition is modest enough to keep $\chi_{\rm B}$ positive,
then this leads to the enhancement of $c_s^2 \propto \chi^{-1}$.
This behavior is what we indeed found in the phase shift approach.

\end{document}